\documentclass[aps,prl,twocolumn,superscriptaddress,showpacs,showkeys]{revtex4-1}

\usepackage{graphicx}% Include figure files

\begin{document}

%\preprint{}

\title{Quasi-two-dimensional Fermi surfaces of the heavy-fermion superconductor Ce$_2$PdIn$_8$}


\author{K. G\"{o}tze}
%\thanks{}
\affiliation{Hochfeld-Magnetlabor Dresden (HLD-EMFL), Helmholtz-Zentrum Dresden-Rossendorf and TU Dresden, D-01314 Dresden, Germany}

\author{J. Klotz}
\affiliation{Hochfeld-Magnetlabor Dresden (HLD-EMFL), Helmholtz-Zentrum Dresden-Rossendorf and TU Dresden, D-01314 Dresden, Germany}

\author{D. Gnida}
\affiliation{Institute of Low Temperature and Structure Research, Polish Academy of Sciences, P.O. Box 1410, PL-50-950 Wroc{\l}aw, Poland}

\author{H. Harima}
\affiliation{Graduate School of Science, Kobe University, Kobe 657-8501, Japan}

\author{D. Aoki}
\affiliation{IMR, Tohoku University, Oarai, Ibaraki 311-1313, Japan}
\affiliation{INAC/SPSMS, CEA Grenoble, 38054 Grenoble, France}

\author{A. Demuer}
\affiliation{Laboratoire National des Champs Magn\'{e}tiques Intenses (LNCMI-EMFL), CNRS, UJF, 38042 Grenoble, France}

\author{S. Elgazzar}
\affiliation{Highly Correlated Matter Research Group, Department of Physics, University of Johannesburg, P.O. Box 524, Auckland Park 2006, South Africa}

\author{J. Wosnitza}
\affiliation{Hochfeld-Magnetlabor Dresden (HLD-EMFL), Helmholtz-Zentrum Dresden-Rossendorf and TU Dresden, D-01314 Dresden, Germany}

\author{D. Kaczorowski}
\affiliation{Institute of Low Temperature and Structure Research, Polish Academy of Sciences, P.O. Box 1410, PL-50-950 Wroc{\l}aw, Poland}

\author{I. Sheikin}
\email[]{ilya.sheikin@lncmi.cnrs.fr}
%\thanks{}
\affiliation{Laboratoire National des Champs Magn\'{e}tiques Intenses (LNCMI-EMFL), CNRS, UJF, 38042 Grenoble, France}


\date{\today}

\begin{abstract}
We report low-temperature de Haas-van Alphen (dHvA) effect measurements in magnetic fields up to 35 T of the heavy-fermion superconductor Ce$_2$PdIn$_8$. The comparison of the experimental results with band-structure calculations implies that the 4$f$ electrons are itinerant rather than localized. The cyclotron masses estimated at high field are only moderately enhanced, 8 and 14 $m_0$, but are substantially larger than the corresponding band masses. The observed angular dependence of the dHvA frequencies suggests quasi-two-dimensional Fermi surfaces in agreement with band-structure calculations. However, the deviation from ideal two dimensionality is larger than in CeCoIn$_5$, with which Ce$_2$PdIn$_8$ bears a lot of similarities. This subtle distinction accounts for the different superconducting critical temperatures of the two compounds.
\end{abstract}

%\keywords{}

\maketitle

The appearance of unconventional superconductivity in the vicinity of a quantum critical point (QCP) is a common trend in Ce-based heavy-fermion (HF) compounds. A more recent and still somewhat controversial issue is the influence of the Fermi-surface (FS) dimensionality on unconventional superconductivity. Indeed, reduced dimensionality of the FS leads to nesting-type magnetic instabilities~\cite{Moriya1990} and thus enhances the superconductivity~\cite{Monthoux1999,Monthoux2003}. The exact knowledge of the FS topology of HF systems is, therefore, essential. In addition, this information allows distinguishing if the $f$ electrons are itinerant or localized, i.e. whether they contribute to the FS or not.

Ce$_2$PdIn$_8$ is a recently discovered HF superconductor with $T_c = 0.7$ K and a non-magnetic ground state~\cite{Kaczorowski2009,Kaczorowski2010}. Non-Fermi-liquid behavior was observed in both macroscopic~\cite{Dong2011,Tran2011,Gnida2012,Tokiwa2011,Matusiak2011} and microscopic~\cite{Fukazawa2012,Tran2012} measurements at low temperature, implying that Ce$_2$PdIn$_8$ is located very close to a QCP. It was further suggested that a two-dimensional (2D) SDW-type QCP is induced by magnetic field near the upper critical field, $H_{c2} \approx$ 2 T~\cite{Tokiwa2011}. Unconventional superconductivity was demonstrated to be due to antiferromagnetic quantum fluctuations~\cite{Hashimoto2013}. These unusual properties are strikingly similar to those of the well-studied HF superconductor CeCoIn$_5$~\cite{Paglione2003,Bianchi2003,Bauer2005,Ronning2006}, which is also located very close to a QCP at ambient pressure. However, the superconducting critical temperature $T_c = 2.3$ K of CeCoIn$_5$~\cite{Petrovic2001} is considerably higher than that of Ce$_2$PdIn$_8$.

Ce$_2$PdIn$_8$ crystallizes into a tetragonal Ho$_2$CoGa$_8$-type crystal structure with space group $P4/mmm$. It belongs to the larger family of Ce$_nT$In$_{3n+2}$ ($T$: transition metal, $n =$ 1, 2, and $\infty$) systems, containing a sequence of $n$ CeIn$_3$ layers intercalated by a $T$In$_2$ layer along the $c$ axis. While cubic CeIn$_3$ ($n =\infty$) is a completely isotropic system, the layered structures with $n =$ 1 and 2 are expected to lead to anisotropic properties and quasi-2D FSs. Indeed, quasi-2D FS sheets were observed in both $n = 1$ systems CeCoIn$_5$~\cite{Settai2001,Hall2001}, CeIrIn$_5$~\cite{Haga2001}, CeRhIn$_5$~\cite{Shishido2002,Hall2002} and the $n = 2$ compound Ce$_2$RhIn$_8$~\cite{Ueda2004,Jiang2015}. The degree of two dimensionality is expected to be larger in monolayer systems with alternating layers of CeIn$_3$ and $T$In$_2$ than in their bilayer counterparts, in which two CeIn$_3$ layers are separated by one $T$In$_2$ layer.

In this paper, we report high-field dHvA measurements of Ce$_2$PdIn$_8$. The observed FS is quasi-2D, however, we find that the more three-dimensional crystal structure of Ce$_2$PdIn$_8$ relative to CeCoIn$_5$ leads to a reduced two dimensionality of the FS topology. We argue that this can explain the difference in the superconducting critical temperatures of the two compounds.

Single crystals were grown by the self-flux method~\cite{Kaczorowski2009}, and we have confirmed by specific-heat measurements that they are not contaminated by CeIn$_3$. The dHvA measurements were performed using a torque cantilever magnetometer mounted in a top-loading dilution refrigerator equipped with a low-temperature rotator. Magnetic fields, $B$, up to 35 T generated by LNCMI-Grenoble resistive magnets were applied at different angles between the [001] and [100] directions.

\begin{figure}[htb]
\includegraphics[width=7.5cm]{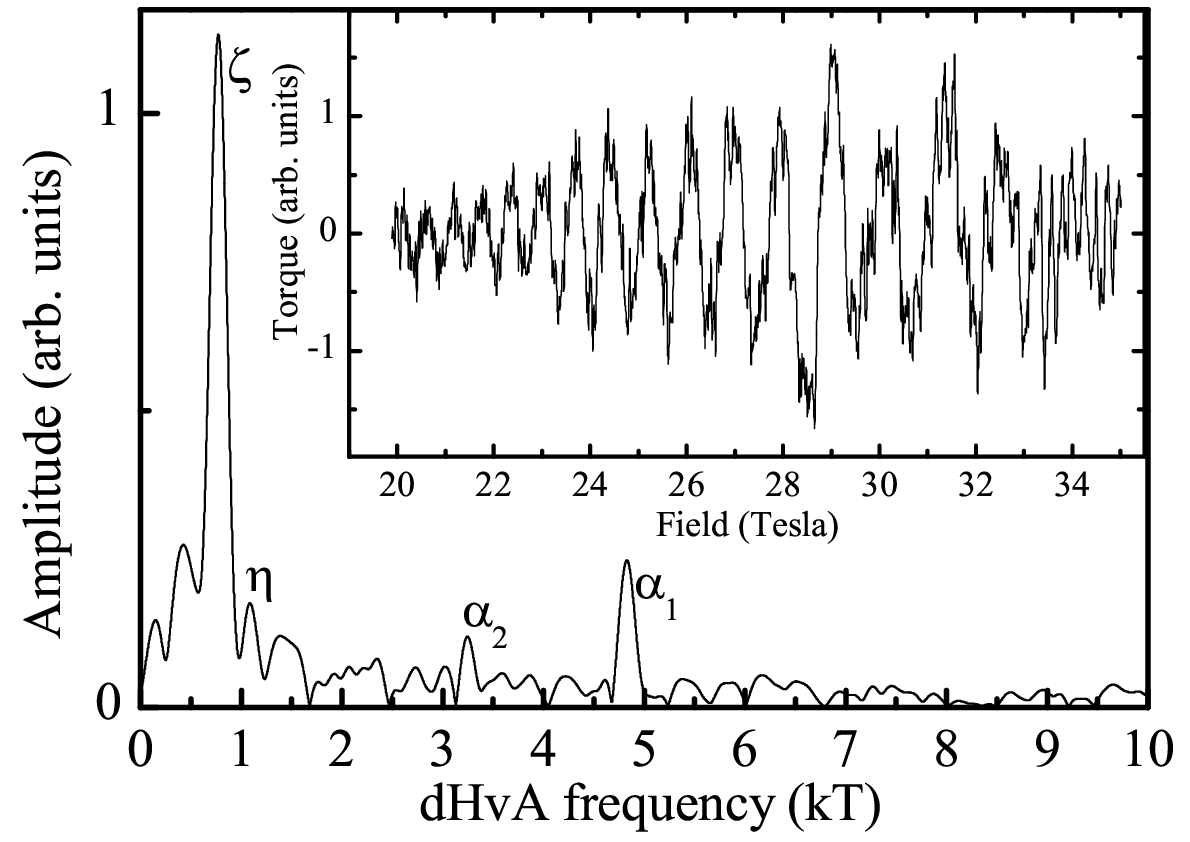}
\caption{\label{dHvA}Fourier spectrum of the high-field dHvA oscillations (inset) in Ce$_2$PdIn$_8$ for magnetic field applied at 4$^\circ$ off the $c$ axis at 30 mK.}
\end{figure}

Figure~\ref{dHvA} shows the oscillatory torque after subtracting a non-oscillating background and the corresponding Fourier transform in Ce$_2$PdIn$_8$. Four fundamental frequencies, denoted $\zeta$, $\eta$, $\alpha_1$, and $\alpha_2$, are observed when $B$ is applied close to the $c$ axis. The oscillations were traced up to 60$^\circ$, where their amplitude decreased below the noise level.

\begin{figure*}[htb]
\includegraphics[width=15cm]{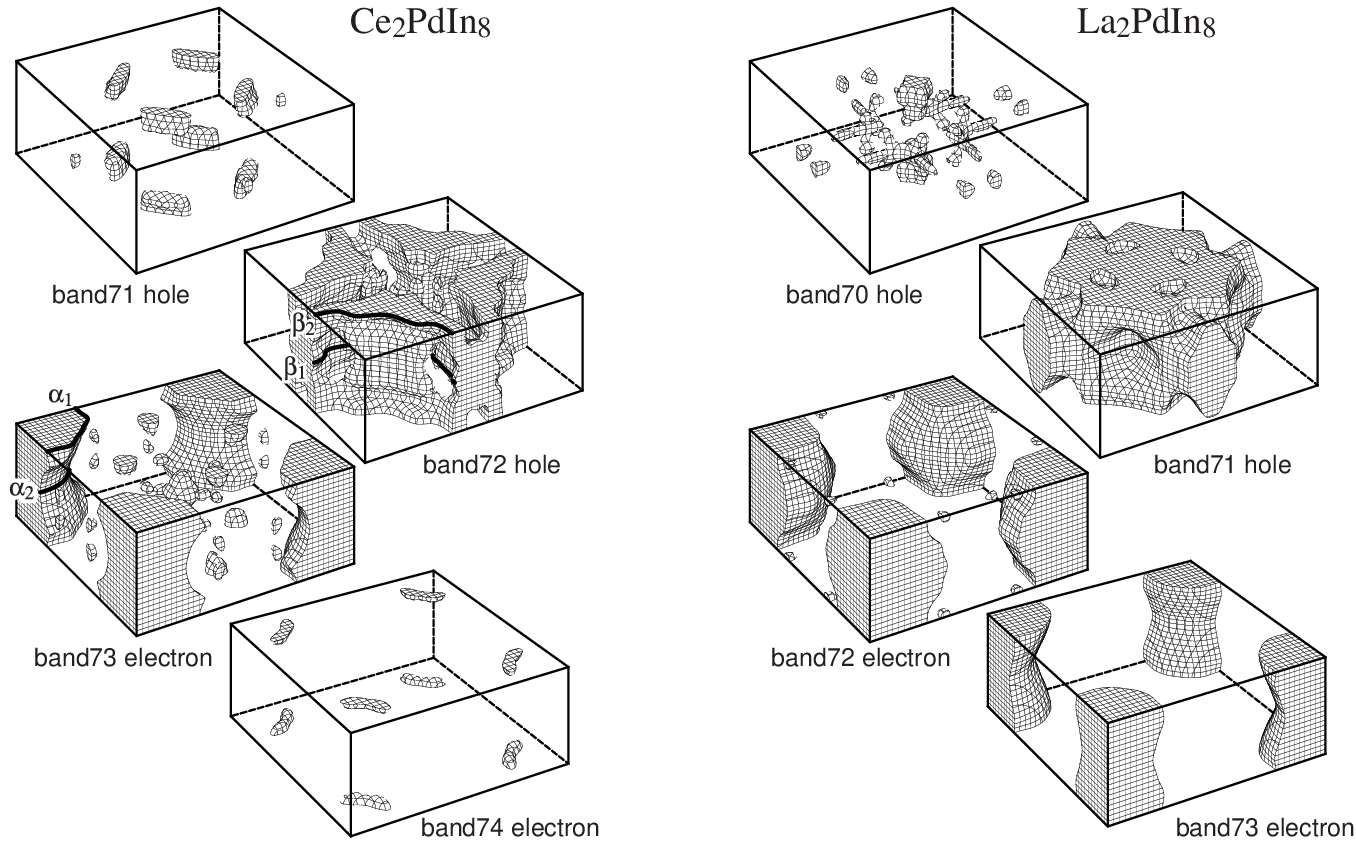}
\caption{\label{FS}Calculated FSs of Ce$_2$PdIn$_8$ (left). Calculated FSs of La$_2$PdIn$_8$ (right) are also shown for comparison.}
\end{figure*}

To figure out whether the $f$ electrons are itinerant or localized in Ce$_2$PdIn$_8$, we performed band-structure calculations for both Ce$_2$PdIn$_8$ and La$_2$PdIn$_8$, the latter corresponding to the Ce compound with localized $f$ electrons. For both compounds, the calculations were carried out using a full potential augmented plane wave method with the local density approximation (LDA) for the exchange-correlation potential. As crystals of La$_2$PdIn$_8$ are currently unavailable, the lattice parameters of Ce$_2$PdIn$_8$ were used for the La$_2$PdIn$_8$ calculations. The resulting FSs are shown in Fig.~\ref{FS}.

Given the layered crystal structure, it is not surprising that some of the calculated FS sheets are quasi-2D in both Ce$_2$PdIn$_8$ and La$_2$PdIn$_8$. The details of the FSs are, however, clearly different. In contrast, the topology of the 4$f$-itinerant FS of CeCoIn$_5$ is similar to the 4$f$-localized FS of LaRhIn$_5$ and CeRhIn$_5$~\cite{Shishido2002,Hall2002}, where the two FSs differ mainly by size. It should be noted that CeCoIn$_5$ is a compensated metal with equal carrier numbers of electrons and holes, while LaCoIn$_5$ is an uncompensated metal. On the other hand, both Ce$_2$PdIn$_8$ and La$_2$PdIn$_8$ are compensated metals. As seen in Fig.~\ref{FS}, the charge-carrier number given by the FS volume in Ce$_2$PdIn$_8$ is about two times smaller than that in La$_2$PdIn$_8$.

\begin{figure*}[htb]
\includegraphics[width=15cm]{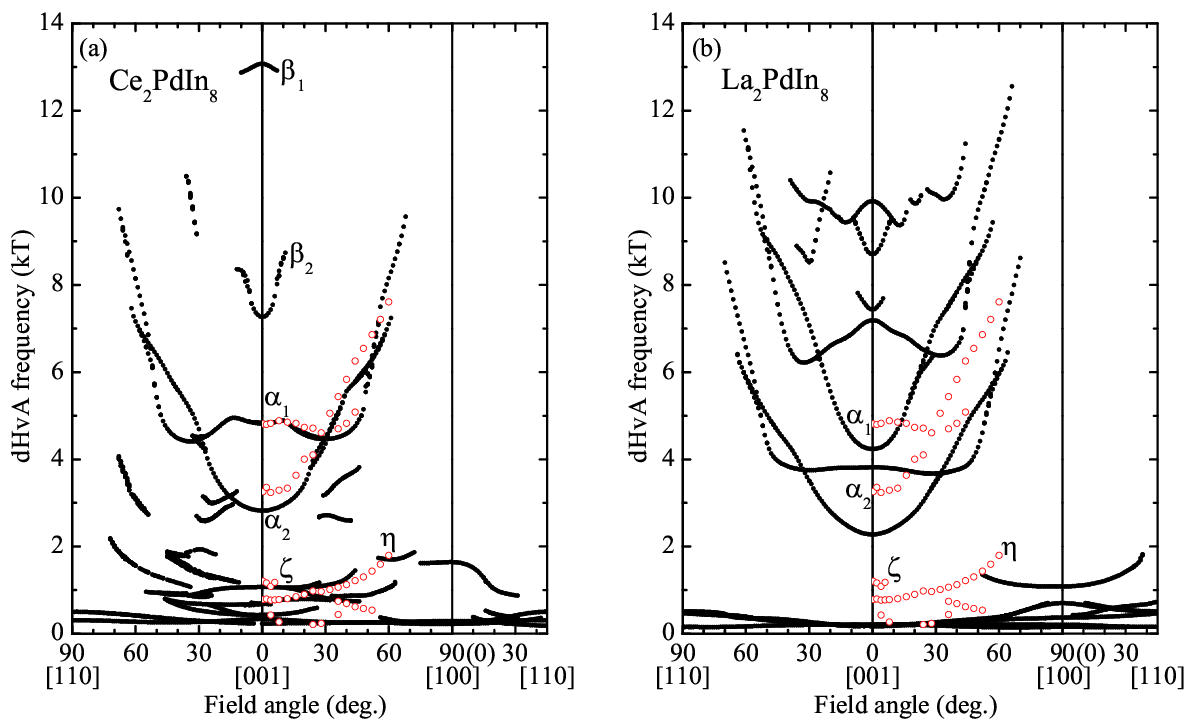}
\caption{\label{AngDep}(color online) Angular dependence of the experimentally observed dHvA frequencies in Ce$_2$PdIn$_8$ (big open circles) is shown together with the results of band-structure calculations (small closed circles) performed for Ce$_2$PdIn$_8$ (a) and La$_2$PdIn$_8$ (b). The latter correspond to Ce$_2$PdIn$_8$ with localized 4$f$ electrons and are shown for comparison. Very low calculated dHvA frequencies that correspond to small FS pockets are not shown for clarity.}
\end{figure*}

\begin{table}[hbt]\centering
\begin{tabular}{c|cc|cc}
\hline
\hline
Branch& \multicolumn{2}{c|}{Experiment}&\multicolumn{2}{c}{Calculation}\\
& $F$ (kT)& $m^{\ast}/m_0$\footnote{The effective masses were measured with magnetic field applied at 4$^\circ$ off the $c$ axis}& $F$ (kT)& $m_b/m_0$\\
\hline
$\gamma$& ---& ---& 0.34& 1.55\\
$\delta$& ---& ---& 0.43& 0.42\\
$\zeta$& 0.78& 8.4$\pm$0.4& 0.69& 1.24\\
$\eta$& 1.2& ---& 1.07& 2.27\\
$\alpha_2$& 3.26& ---& 2.82& 0.81\\
$\alpha_1$& 4.82& 14$\pm$1& 4.82& 2.12\\
$\beta_2$& ---& ---& 7.27& 2.81\\
$\beta_1$& ---& ---& 13.08& 5.2\\
\hline
\hline
\end{tabular}
%\vspace{3mm}
\caption{\label{tab:dHvA_parameters}Experimental and calculated dHvA frequencies and effective masses in Ce$_2$PdIn$_8$ for magnetic field along the $c$ axis.}
\end{table}

Figure~\ref{AngDep}(a) shows the experimentally observed angular dependence of the dHvA frequencies in Ce$_2$PdIn$_8$ together with the results of band-structure calculations based on the 4$f$-itinerant band model. Experimental and calculated frequencies and effective masses are also shown in Table~\ref{tab:dHvA_parameters}. The agreement between the experimentally observed $\alpha$ branches and those of the calculations is excellent. Not only the angular dependences are the same, but even the absolute values agree very well. This implies that both the topology and the size of the calculated FS sheet reproduce the experimental results exceptionally well. The $\alpha$ branches correspond to the quasi-2D FS of the band 73, as shown in Fig.~\ref{FS}. Regarding the measured lower ($<2$ kT) dHvA frequencies, there is also a very good agreement between the experimental and calculated branches. These branches originate mostly from rather isotropic parts of the FS in band 72. The calculated $\beta$ branches originating from complicated sheets in band 72 were, however, not observed in the experiment. This is probably caused by a strongly enhanced effective mass or an unfavorable curvature factor for detecting the dHvA signal. For comparison, in Fig.~\ref{AngDep}(b) we plot the experimental results obtained in Ce$_2$PdIn$_8$ together with band-structure calculations for La$_2$PdIn$_8$. In this case, the two are obviously at odds with each other. In particular, only one quasi-2D FS was observed in the experiment, while the calculations predict two of them for La$_2$PdIn$_8$, which should be easy to detect.

The comparison of the experimentally observed dHvA frequencies with the results of the LDA band-structure calculations thus gives clear evidence for a quasi-2D FS with itinerant $f$ electrons in Ce$_2$PdIn$_8$. The same conclusion was drawn for CeCoIn$_5$~\cite{Settai2001,Hall2001}.

We, alternatively, calculated the band structure using the local spin density approximation with the relativistic version of the full-potential local orbital method~\cite{Koepernik1999} for Ce$_2$PdIn$_8$. This also suggests quasi-2D FS, but the agreement with the experimental results is not as good.

The effective masses shown in Table~\ref{tab:dHvA_parameters} were determined by fitting the temperature dependence of the oscillatory amplitude by the standard Lifshitz-Kosevich formula~\cite{Shoenberg2009}. This was done for the magnetic field applied at 4$^\circ$ off the $c$ axis. Due to the small amplitudes of the oscillations the field range from 28 to 34.5 T was used for the analysis. Even for such high fields, the effective masses of only two branches, $\zeta$ and $\alpha_1$, could be reliably determined. The obtained values are 8.4$\pm$0.4 and 14$\pm$1 $m_0$, respectively. The effective mass of the $\alpha_1$ branch corresponding to the quasi-2D sheet of the FS is comparable to the values, 8 - 18 $m_0$, reported for the quasi-2D FS of CeCoIn$_5$~\cite{Settai2001,Hall2001,Polyakov2012}. This implies a similar degree of hybridization between the $f$ and conduction electrons. The detected effective masses are, however, by far too small to account for the huge value of the electronic specific heat coefficient, $\gamma$, of the order of 1 J/K$^2$mol just above the superconducting transition~\cite{Kaczorowski2009,Kaczorowski2010,Tokiwa2011}. Presumably, the effective masses of the $\beta$ branches, which are not observed here, are strongly enhanced. Indeed, already the calculated band masses of the $\beta$ branches are higher than those of the other branches (see Table~\ref{tab:dHvA_parameters}). On the other hand, the Sommerfeld coefficient of Ce$_2$PdIn$_8$ is similar to that of CeCoIn$_5$~\cite{Petrovic2001}, where the effective masses were reported to strongly decrease with magnetic field~\cite{Settai2001}. While the observed dHvA oscillations in Ce$_2$PdIn$_8$ are not strong enough to perform the field-dependent analysis of the effective masses, they can also be expected to decrease with magnetic field. This assumption is supported by the experimentally observed field dependence of the $T^2$-coefficient in the resistivity~\cite{Dong2011} and of the Sommerfeld coefficient of the specific heat~\cite{Tokiwa2011} above the upper critical field.

\begin{figure}[htb]
\includegraphics[width=6cm]{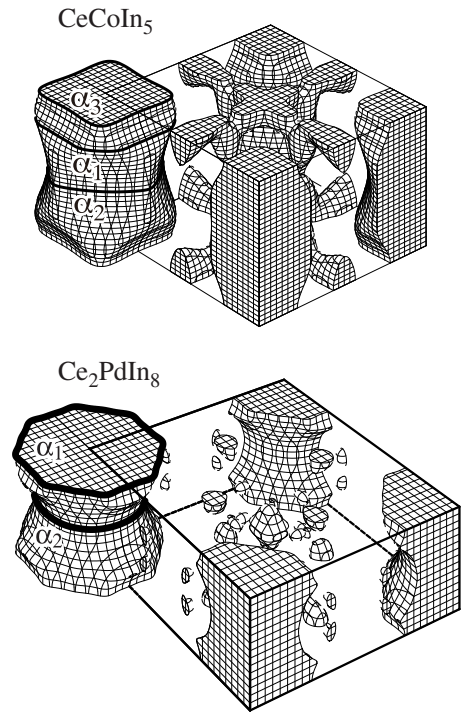}
\caption{\label{FS_comparison}Comparison of the calculated quasi-2D FSs of CeCoIn$_5$ and Ce$_2$PdIn$_8$.}
\end{figure}

As shown in Fig.~\ref{FS_comparison}, the major FS sheets of both Ce$_2$PdIn$_8$ and CeCoIn$_5$ are quasi-2D corrugated cylinders extending along the [001] direction. As many of the physical properties of HF materials strongly depend on the FS dimensionality, the key question here is which FS is more 2D, i.e. which amplitude of the corrugation is smaller. For both Ce$_2$PdIn$_8$ and CeCoIn$_5$, the FSs experimentally determined through dHvA measurements are in excellent agreement with calculated ones. This is, however, not always the case. That is why we introduce a quantitative criterion of a quasi-2D FS deviation from an ideal cylinder: $\Delta = (S_{max} - S_{min})/\overline{S}$, where $S_{max}$ and $S_{min}$ are the maximum and minimum extremal cross-sections respectively, and $\overline{S}$ is the average cross-section of the warped cylinder. For an ideal cylinder, $\Delta = 0$. Since the extremal cross-sections, $S_i$, of the FS are proportional to the dHvA frequencies, $F_i$, $S$ can be replaced by $F$ measured with field along [001] to determine $\Delta$ experimentally. In Ce$_2$PdIn$_8$, with the two frequencies, $\alpha_1$ and $\alpha_2$, listed in Table~\ref{tab:dHvA_parameters}, this yields $\Delta = 0.386$. The dHvA effect measurements in CeCoIn$_5$ revealed three extremal cross-section of the quasi-2D FS~\cite{Settai2001,Hall2001,Polyakov2012}. The reported values of the dHvA frequencies are slightly different, yielding the average value of $\Delta = 0.221$. This implies that the deviation from the ideal 2D FS is much smaller in CeCoIn$_5$ than in Ce$_2$PdIn$_8$. This is expected as well from the more 3D crystal structure of Ce$_2$PdIn$_8$ as compared to CeCoIn$_5$.

The corresponding larger anisotropy of CeCoIn$_5$ as compared to Ce$_2$PdIn$_8$ accounts for the higher superconducting critical temperature of CeCoIn$_5$. In fact, 2.3 K in CeCoIn$_5$ is the highest $T_c$ among all the known Ce-based HF materials. Remarkably, the FS of CeCoIn$_5$ is the most 2D like as compared to its Ir and Rh analogs~\cite{Hall2001}. The FS of  CeRhIn$_5$, however, changes at its critical pressure $P_c \simeq 2.4$ GPa~\cite{Shishido2005}, and the reported dHvA frequencies yield $\Delta = 0.17$ above $P_c$. This value is similar to that of CeCoIn$_5$ at ambient pressure. In CeRhIn$_5$, superconductivity emerges around $P_c$, where $T_c = 2.1$ K~\cite{Hegger2000}, a value close to that of CeCoIn$_5$. Regarding CeIrIn$_5$, experimentally observed dHvA frequencies~\cite{Haga2001} result in $\Delta = 0.269$, a value in between those for CeCoIn$_5$ and Ce$_2$PdIn$_8$. However, $T_c = 0.4$ K of CeIrIn$_5$~\cite{Petrovic2001a} can not be compared directly to the critical temperatures of CeCoIn$_5$ and Ce$_2$PdIn$_8$, as CeIrIn$_5$ is located further away from a QCP~\cite{Matsumoto2010}. When CeIrIn$_5$ is tuned to a QCP by Rh substitution, $T_c$ increases to about 1 K~\cite{Zheng2004}, and is likely to be reduced due to disorder as compared to pure compounds. Consistently with $\Delta$, this value also falls in between those for CeCoIn$_5$ and Ce$_2$PdIn$_8$. Unfortunately, there is currently no information about the FS of Rh-substituted CeIrIn$_5$. While the FS dimensionality is not the only factor that determines $T_c$ in HF superconductors, it is certainly a significant one. Indeed, $T_c = 18.5$ K was reported for PuCoGa$_5$~\cite{Sarrao2002}, which is the highest among those yet observed in $f$-electron materials. Remarkably, the calculated FS of PuCoGa$_5$ consists of three corrugated cylinders~\cite{Maehira2003}, although the degree of corrugation is relatively high with $\Delta$ being 0.448, 0.359 and 0.66 respectively. However, the results of these calculations are still to be confirmed experimentally.

In summary, our high-field dHvA investigation of Ce$_2$PdIn$_8$ combined with band-structure calculations evidence the existence of s quasi-2D FS with itinerant $f$-electrons in this compound. The comparison of the FS topology of Ce$_2$PdIn$_8$ and CeCoIn$_5$ implies that the FS of the latter compound is much closer to an ideal cylinder characteristic for a 2D case. The difference in the FS dimensionality accounts for different superconducting critical temperatures of the two compounds, which are both located in close vicinity to a QCP and have a similar degree of hybridization between the 4$f$ and conduction electrons. It would be interesting to apply the quantitative criterion of the FS two dimensionality we introduced here to other HF materials with quasi-2D FS. In particular, the criterion can be used to verify the theoretical prediction about the influence of the FS dimensionality on the type of quantum criticality in HF compounds~\cite{Si2014,Si2006,Si2010a,Custers2012}. Another interesting question is whether magnetic fields themselves affect the FS dimensionality in Ce$_2$PdIn$_8$ in particular and other quasi-2D HF materials in general. Zero-field ARPES measurements in Ce$_2$PdIn$_8$ would be very useful to address this issue.

\begin{acknowledgments}
We are grateful to T. Maehira for sharing with us the details of the band-structure calculations in PuCoGa$_5$. KG acknowledges support from the DFG within GRK 1621. We acknowledge the support of the HLD-HZDR and the LNCMI-CNRS, members of the European Magnetic Field Laboratory (EMFL).
\end{acknowledgments}

\bibliography{Ce2PdIn8}

\end{document}